\documentstyle[aps,prl,multicol,epsf]{revtex}
\begin{document} 
\draft 
\title{Are Directed Waves Multifractal?} 
\author{Yi-Kuo Yu$^{1,2}$ and H. Mathur$^{1}$} 
\address{$^{1}$Department of Physics,
Case Western Reserve University, Cleveland, Ohio
44106-7079} 
\address{$^{2}$Department of Physics, Florida Atlantic University,
777 Glades Road, Boca Raton, FL 33431}

\date{\today} 
\maketitle


\begin{abstract}
Wave propagation is studied in a sufficiently anisotropic
random medium that backscattering along one direction can
be neglected. A Fokker-Planck equation is derived the solution
to which would provide a complete statistical description 
of such directed waves. The Fokker-Planck equation is
mapped onto an su(1,1) ferromagnet and its symmetries are
identified. Using the symmetries asymptotic wave 
function distributions are computed and used to show that
directed wave functions fill space uniformly and do {\em 
not} have multifractal character.

\end{abstract} 

\begin{multicols}{2} 

Much progress has been achieved in the study of 
wave propagation in random media by the infusion
of scaling ideas from statistical mechanics and
critical phenomena. Dirty electronic systems are
the best studied example \cite{lee,huckestein}. 
They are now known to
exhibit a variety of phases and transitions, many
of which remain poorly understood. 
Recently there has been interest in the application
of multifractal analysis to these problems \cite{huckestein}. 
Fractals are complex geometric forms that exhibit self-similarity
under magnification \cite{mandelbrot}. Multifractals are still more
complex objects which need an infinite set of exponents
to characterise their self-similar scaling rather than
the single exponent that suffices for an ordinary fractal 
\cite{kadanoff}.
These ideas will be expressed precisely below; no 
previous background in multifractal analysis is needed
to read this paper.

In statistical mechanics the typical configuration
of a disordered spin-system
at a critical point is complicated and for certain models
can be shown to exhibit multifractal scaling \cite{ludwig}.
The purpose of such an analysis is to provide a 
characterisation of the critical point in terms of
multifractal exponents that supplement the conventional
universal exponents of critical phenomena \cite{ma}. Dirty 
electronic
wave functions have also been studied from this perspective.
For example, extensive numerical simulations indicate that
the wave functions of a quantum Hall system are multifractal
at the critical point that separates two Hall resistance
plateaus \cite{huckestein}. 
Calculation of multifractal exponents requires
detailed statistical information about the  
wave functions and analytic results are
therefore difficult to obtain. With few exceptions
\cite{ludwig2}, multifractal analysis for dirty 
electronic systems is therefore carried out via
numerical simulation.
In this Letter we report multifractal analysis of
directed wave propagation through a random
medium. This problem has been investigated by
the authors of ref \cite{multifractal} who claimed numerical evidence
of multifractal scaling.

The directed wave model is very rich in physical applications.
It applies whenever waves propagate in a sufficiently anisotropic
medium that backscattering in one direction can be neglected
\cite{feng,feng2}. 
An experimental realization \cite{stormer,chaikin,druist}
that has attracted much attention recently 
\cite{chalker,balents,kim,matthew,me,yikuo,ilya} (and refs
therein) is the surface of a quantum Hall multilayer.
Studies of directed waves have a long history due to the
equivalence of this problem to time-dependent wave propagation
in a noisy environment \cite{russian,multifractal}. Moreover
a recent model for stress propagation in granular materials
\cite{sue} is identical to lattice models of directed
waves \cite{saul}. Still another motivation for studying the
directed wave model has been its formal similarity 
to the directed polymer problem \cite{stanley} 
(for an illuminating discussion of this viewpoint
see, for example, ref \cite{saul}).

Previous analytic work on directed waves has focussed on calculation of
low-order correlations of the wave functions (needed, for example,
to calculate
observables such as the disorder averaged conductance
of a quantum Hall multilayer). Here
it is neccessary to analyse the full probability distribution
of the wavefunctions. Our method is to derive a Fokker-Planck
equation that governs the evolution of 
the probability distribution. This
equation has an su(1,1) symmetry that can be exploited to
obtain the asymptotic probability distribution which
reveals that directed waves fill space quite
uniformly and therefore do not change appearance under 
magnification. Thus, contrary to  
previous numerical work, 
we are able to show exactly
that directed wavefunctions do {\em not} have a multifractal 
character. The asymptotic wave function distribution (eq 3 below)
is the directed wave analogue of a result derived several
years ago for directed polymers in one-dimension by Huse
{\em et al.} \cite{huse}.

We shall take directed waves at a fixed frequency to 
evolve according to 
(see for example, \cite{matthew,me})
\begin{equation}
-i \frac{ \partial }{ \partial x } \psi_{n} ( x )
+ t_{n} ( x ) \psi_{n+1} ( x ) 
+ t_{n-1}^{*} ( x ) \psi_{n-1} ( x )
= 0.
\end{equation}
Backscattering is neglected along the $ x-$direction. The
transverse direction is taken to be discrete and the system
is assumed to be of finite size, $N $, in this direction
(with periodic boundary conditions). The anisotropic
and directional
nature of eq (1) is reflected in the fact that it is of
first order in the $x-$direction whereas it is of second order
in the transverse direction. Disorder is incorporated by
taking the hopping terms $t_{n}(x)$ 
to be random with statistics given below. 
Other models have been considered in the literature. They
are either equivalent to eq (1) or are believed to have the
same qualitative behaviour\cite{onsite}.
The problem
posed is the following: The
wavefunction is specified for a fixed value of $x$; it is
then evolved in the $x-$direction according to eq (1). For
definiteness, it may be supposed that the wavefunction is
localized at $n=0$ at $x=0$. Thus $ \psi_{n}(x=0) = \delta_{n,0} $.
Each disorder realization will produce a different wave
function at a larger value of $x > 0$. We are interested in
$P(a_{n},x)$, the probability density that at a fixed $x > 0$ 
the wave function is $ \psi_{n} (x) = a_{n} $.

Our problem resembles the problem of Brownian motion,
in which a heavily damped particle is subject to a
noisy environment. Standard methods exist in the theory
of Brownian motion to convert the stochastic equation
of motion (called the Langevin equation) into the
corresponding Fokker-Planck equation which governs the
time evolution of the probability density of the particle
position \cite{reif}. 
These methods may be applied to obtain the
Fokker-Planck equation corresponding to eq (1)
\begin{eqnarray}
&&-\partial P (a_n,x)/\partial x ={\cal H} P;\nonumber \\
&&{\cal H}= D\sum_{\alpha,m} a_m^\alpha {\partial \over \partial a_m^\alpha}
\nonumber \\
&&\ \ \ \ -{D\over 4}\sum_{\alpha,\beta,m}\big[(a_{m+1}^\alpha)^2
{\partial^2 \over \partial {a_m^\beta}^2} + (a_m^\alpha)^2
{\partial^2 \over \partial {a_{m+1}^\beta}^2}\big]\nonumber  \\
&&\ \ \ \ +{D \over 2}\sum_{\alpha,\beta,m} 
a_m^\alpha {\partial \over \partial
a_m^\alpha} a_{m+1}^\beta {\partial \over \partial a_{m+1}^\beta}
\nonumber \\
&&\ \ \ \ -{D \over 2}\sum_{\alpha,\beta,\rho,\nu,m}
\epsilon_{\alpha,\beta} \epsilon_{\rho,\nu}
 a_m^\alpha {\partial \over \partial a_m^\beta} 
 a_{m+1}^\rho {\partial \over \partial
 a_{m+1}^\nu} 
\end{eqnarray}
where the Greek letters represent either $1$ or $2$.
Here we have refined our notation to write $a^1=$ real
part of the wave function and $a^{2}=$ imaginary part.
Likewise $t^{1}$ and $t^{2}$ denote the real and
imaginary parts of the hopping term. In deriving
eq (2) it was assumed that the hopping is a
Gaussian white noise process with zero mean
and variance given by $ [ t^{\alpha}_{n} (x) t^{\beta}_{m} (x') ]_{{\rm
ens}} = (D/2) \delta_{\alpha,\beta} \delta_{n,m} \delta ( x - x' ) $.
$ [ \ldots ]_{{\rm ens}} $ denotes an average over 
the ensemble of disorder realizations. Given an 
initial wavefunction, $ \psi_{n}(x=0) $,
the Fokker Planck eq (2) in principle allows
calculation of the complete probability distribution
of the wave function, $ P(a,x)$ for larger values of
$ x $. In practice, this may appear difficult because
eq (2) is a partial differential equation in a large
number (2N+1) of variables. However it will be
seen that eq (2) has a number of helpful symmetries.

First note that stationary solutions to the Fokker
Planck equation must be radial. It is easy to 
verify by substitution that if $ P $ is radial
(a function only of $ \sum_{n,\alpha} ( a_{n}^{\alpha} )^{2} $
and possibly $ x $) then $ \partial P / \partial x = 0 $
showing that radial solutions are stationary. The
converse, that stationary solutions {\em must}
be radial, is also true and will be shown below.
It is a consequence of an su(1,1) symmetry of the
Fokker-Planck equation to which we shall return.

A second useful property of eq (2) follows from
probability conservation. By direct substitution
into the wave eq (1), it can be shown that the
total probability $ \sum_{n, \alpha} ( \psi^{\alpha}_{n} )^{2} $
does not change with $ x $ for any disorder realization.
Consequently if we solve the Fokker-Planck equation
subject to the initial condition that the system
begins with some definite normalized wave function,
$ P $ must live on the unit sphere in $a-$space. In other
words, $P $ vanishes unless $ \sum_{n, \alpha} ( a_{n}^{\alpha} )^{2}
= 1$.

The only stationary distribution consistent with 
probability conservation is a uniform distribution
on the unit sphere in $ a-$space. Thus the large $x$
asymptotic distribution is uniquely determined to be
\begin{equation}
P(a, \infty) = \frac{1}{A} \delta ( \sum_{n, \alpha} ( a^{\alpha}_{n} )^{2}
- 1 ).
\end{equation}
Here $ A $ is a constant fixed by the normalization condition
$ \int \prod_{n, \alpha} d a^{\alpha}_{n} P( a, \infty ) = 1$.
The physical content of eq (3) is that directed wave functions
are completely randomized after propagating a large distance;
however they remain normalized. Precisely the same distribution
arises in connection with the distribution of matrix elements
of correlated random matrices \cite{mehta}; thus techniques
for averaging over such distributions are well developed.

Multifractal characterization of the directed wave functions
begins with calculation of the exponents $\zeta (q)$
which are defined via \cite{multifractal}
\begin{equation}
[ \sum_n(\sum_\alpha | \psi^{\alpha}_{n} |^2)^q]_{{\rm ens}}
\sim N^{\zeta(q)}.
\end{equation}
The generalized fractal dimensions $D(q)$ are related to these
exponents by $ \zeta(q) \equiv (1-q) D(q)$.  
For $ q =2 $, the left hand side of eq (4) is the
inverse of the
participation ratio which has long been studied in
connection with disordered electronic systems \cite{thouless}.
Roughly it measures the number of sites on which
the wave function has a substantial weight. Eq (4)
shows that $ \zeta(q) $ characterizes the growth of 
generalized inverse participation ratios with system size.
Evidently $\zeta(0) = 1$ and $ \zeta(1) = 0 $ (by normalization).
To calculate the multifractal dimensions for other values
of $ q $, it is neccessary to average the inverse
participation ratios over the wave function distribution
in eq (3).

Evaluation of the necessary integrals is facilitated
by a trick given in ref \cite{mehta}. For illustration
consider the normalization integral $ A = \int d a 
\delta (a^{2} - 1)$. Here for brevity $ \prod_{n, \alpha} d 
a^{\alpha}_{n} $ is written as $ d a $;
$\sum_{n, \alpha} (a^{\alpha}_{n})^{2} $ as $ a^{2} $;
and indices are suppressed everywhere. Rescale by
introducing $ b = a \sqrt{r} $ and find $ r^{N-1} A
= \int d b \delta (b^{2} - r ) $. If both sides are
multiplied by $ e^{-r}$ and integrated over the rescaling
factor $ r $ from
zero to infinity, the remaining $ b $ integrals become
straightforward gaussians and the final 
result is $ A = \pi^{N}/\Gamma(N) $.
The other integrals can be evaluated in a similar 
fashion to yield
\begin{equation}
[ \sum_n(\sum_\alpha | \psi^{\alpha}_{n} |^2)^q]_{{\rm ens}}=
{\Gamma (N+1) \Gamma (q+1)\over \Gamma (N+q)} \simeq \Gamma(q + 1)
N^{1 - q}.
\end{equation}
The last approximate equality holds for large $N$. Comparing
eq (4) and (5) we see that $ \zeta(q) = 1-q$ (which implies
$D(q)=1$). This is precisely
the scaling expected of an object that is
{\em not} multifractal.

To make contact with previous numerical work \cite{multifractal}
it is neccessary to calculate the $g(\beta)$ spectrum which is defined
as follows \cite{kadanoff,multifractal}: 
Let $w$ be the weight (modulus square
of the wave function) on a particular site. $\beta$ is a logarithmic
measure of the weight defined as $\beta \equiv \ln w/ \ln N$. Assemble
a histogram of $\beta$ values by drawing from the ensemble of disorder
realisations and let $\Pi_{N} (\beta') d \beta $ represent the frequency
with which $\beta$ lies between $ \beta' $ and $\beta' + d \beta$ in a
system with $N$ sites. For large $N$ it is expected that $ \Pi_{N} (
\beta ) \sim N^{g(\beta)}$ which defines $ g(\beta)$. 
To calculate
$g(\beta)$ it is helpful to first evaluate $P(w)dw$, the probability
that the weight on a particular site lies between $w$ and $w + d w$,
obtained by  integrating the distribution in eq (3) over
all sites except one. For large $N$ the result is $P(w) = N \exp( - N w )$.
Straightforward substitution into the definition above then yields
\begin{equation}
g(\beta) = 2 + \beta 
\end{equation} 
for $ \beta < -1$. Note $ g(\beta) > 0 $
over the range $ -2 < \beta < -1$. 

In ref \cite{multifractal} 
the $g(\beta)$ spectrum is calculated numerically
and found to have support on the interval $ -1.97 <\beta< -0.88$.
The authors of ref \cite{multifractal} 
assert (incorrectly) that if $g(\beta)$
has support over a range away from $\beta = -1$, it neccessarily
implies multifractal scaling. This is the primary basis for their
claim of multifractal scaling as the actual deviation from ordinary
scaling they infer is very small: $ \zeta (q) $ (numerical) $=
1 - q + 3.6 \times 10^{-3} q^2 $.
The asymptotic distribution derived here, eq (3),
explicitly shows both ordinary scaling and a spectrum, $g(\beta)$,
with an extended support away from $\beta = -1$. Moreover the form
of $g (\beta)$ we obtain, eq (6), provides a good fit to the
numerical data (fig 3 of ref \cite{multifractal}). Hence we believe
that the numerical data of ref \cite{multifractal} are in fact
consistent with our conclusion that directed waves are not 
multifractal.

To complete the analysis we must now return to the Fokker-Planck
equation and show that all stationary solutions are radial.
Note that eq (2) has the appearance of an imaginary time
Schr\"{o}dinger equation. From this point of view $P(a,x)$ is
the wave function and we are interested in the multiplet of
eigenfunctions of the ``Hamiltonian'', 
${\cal H}$, with eigenvalue zero. It is useful to 
express ${\cal H}$ in 
second quantized language by introducing harmonic oscillator
ladder operators
\begin{equation}
b^{\alpha}_{n} = \frac{1}{\sqrt{2}} \left( a^{\alpha}_{n}
+ \frac{ \partial }{\partial a^{\alpha}_{n} } \right);
\hspace{2mm}
(b^{\alpha}_{n} )^{\dagger} = \frac{1}{\sqrt{2}} \left( a^{\alpha}_{n}
- \frac{ \partial }{\partial a^{\alpha}_{n} } \right);
\end{equation}
which obey the bosonic commutation relations
$[ b^{\alpha}_{n}, b^{\beta \dagger}_{m} ] = \delta_{\alpha \beta}
\delta_{n m}$ and $ [ b^{\alpha}_{n}, b^{\beta}_{m} ] = 0 $.
In this language the Hamiltonian is given by
\begin{eqnarray}
{\cal H}&=& {D\over 2}\sum_n \biggr[2 \sum_\alpha {b^{\alpha}_{n}}^\dagger
 b^{\alpha}_{n}+\sum_{\alpha,\beta} {b^{\alpha}_{n}}^\dagger b^{\alpha}_{n}
{b^{\beta}_{n+1}}^\dagger  b^{\beta}_{n+1} \nonumber \\
&&\ \ \ \ \ \ -{1\over 2}\sum_{\alpha,\beta} ({b^{\alpha}_{n}}^\dagger
 {b^{\alpha}_{n}}^\dagger b^{\beta}_{n+1} b^{\beta}_{n+1} +
b^{\alpha}_{n} b^{\alpha}_{n} {b^{\beta}_{n+1}}^\dagger 
{b^{\beta}_{n+1}}^\dagger )\nonumber \\
&&\ \ \ \ \ \ +\sum_{\alpha,\beta,\rho,\nu} \epsilon_{\alpha,\beta}
\epsilon_{\rho,\nu} {b^{\alpha}_{n}}^\dagger b^{\beta}_{n}
{b^{\rho}_{n+1}}^\dagger b^{\nu}_{n+1}\biggr]
\end{eqnarray}
Thus the Fokker-Planck equation is seen to
be equivalent to an interacting boson problem. 

The symmetry of the Hamiltonian is revealed by considering
the algebra of the bosonic bilinear terms out of which
it is built. Define
$K^{+}_{n}  \equiv   {1\over 2} \sum_\alpha 
{b^{\alpha}_{n}}^\dagger {b^{\alpha}_{n}}^\dagger $,
$ K^{-}_{n}  \equiv {1\over 2} \sum_\alpha 
b^{\alpha}_{n} b^{\alpha}_{n} $,
$ K^{z}_{n}  \equiv  {1\over 2}( {b^1_{n}}^\dagger b^1_{n}
+ b^2_{n} {b^2_{n}}^\dagger ) $ and
$ M_{n}  \equiv  i({b^1_{n}}^\dagger b^2_{n}-{b^2_{n}}^\dagger b^1_{n})
$. These operators obey
\begin{equation}
[ K^{+}, K^{-} ] = - 2 K^{z}; \hspace{2mm}
[ K^{z}, K^{\pm} ] = \pm K^{\pm}
\end{equation}
and $ M $ commutes with all the $K$'s. In writing
eq (9), the site indices have
been suppressed for clarity. These commutation
relations apply only if the indices coincide.
Operators with distinct indices all commute. This
algebra is reminiscent of the angular momentum 
algebra$-$the notation was chosen to highlight
the similarity. However there is a crucial sign
difference in the $[ K^{+}, K^{-} ]$ commutator.
Thus the $ \vec{K} $ operators actually
obey the su(1,1) algebra rather than su(2).
su(1,1) is a well studied classical Lie algebra
sometimes called the {\em hyperbolic} angular
momentum algebra in the physics literature when
it is discussed in connection with Schwinger's
coupled-boson description of angular momentum
\cite{Schwinger}.

The Hamiltonian may be rewritten in terms of these
operators as an su(1,1) ferromagnet (a generalization
of the ordinary Heisenberg ferromagnet in which su(1,1)
operators replace their spin counterparts) \cite{footnote}:
\begin{eqnarray}
{\cal H}&=& D\sum_n \big[ 2K_n^z K_{n+1}^z - (K_n^+ K_{n+1}^-
+ K_n^- K_{n+1}^+ )\nonumber \\
&&\ \ \ \ \ \ \ \ \ \  - {1\over 2}(M_n M_{n+1}+1)\big]
\end{eqnarray}
It is now easy to verify that the Hamiltonian is invariant
under su(1,1) rotations:
\begin{equation}
[ {\cal H}, \vec{K}_{{\rm tot}} ] = 0
\end{equation}
where $ \vec{K}_{{\rm tot}} \equiv \sum_{n} \vec{K}_{n}$

The complete set of eigenstates with eigenvalue zero may
now be found. On account of the symmetry expressed in eq (11),
this multiplet should form an irreducible representation of
the su(1,1) algebra. Note that the boson vacuum is a zero energy
eigenstate of the Hamiltonian. This is evident from inspection
of eq (8). In addition, the vacuum is annihilated by $ K^{-}_{{\rm
tot}} $. Thus the vacuum is the lowest weight state in the 
multiplet; the complete infinite set may be obtained by repeatedly
applying $K^{+}_{{\rm tot}} $ to the vacuum \cite{Schwinger}.

To see that each member of the 
multiplet $ (K^{+}_{{\rm tot}} )^{l} |0> $ with
$ l = 0, 1, 2, \ldots $ corresponds to a radial function
it is neccessary to translate back into first quantized 
language. The ground state of a harmonic oscillator has
a gaussian wave function; hence the boson vacuum is given
by
\begin{equation}
|0> \rightarrow \exp - \frac{1}{2} \sum_{n, \alpha} 
(a^{\alpha}_{n})^{2}
\end{equation}
$-$a radial function. $K^{+}_{{\rm tot}} $ is given by
\begin{equation}
K^{+}_{{\rm tot}} \rightarrow \frac{1}{4}
\sum_{n \alpha} \left[ (a^{\alpha}_{n})^{2}
+ \frac{ \partial^{2} }{ \partial a^{\alpha 2}_{n} }
- 2 a^{\alpha}_{n} \frac{ \partial }{ \partial a^{\alpha}_{n} }
- 1 \right]
\end{equation}
It is easy to verify that the action of $K^{+}_{{\rm tot}} $
on {\em any} radial function will yield another radial
function. Thus we have shown that the vacuum is radial
as is any state built out of it by repeated application
of the raising operator $K^{+}_{{\rm tot}}$; in other
words, a stationary solution must be radial.

Finally, note that although in this paper we have
focussed on stationary solutions, it may be possible
to give a rather complete analysis of the directed
wave problem due to the equivalence demonstrated
here to a relatively
simple model$-$an su(1,1) ferromagnet. 

In summary the main results of this paper are:
derivation of a Fokker-Planck equation that governs
the wave function distribution for directed waves;
mapping of this equation onto an su(1,1) spin-chain
which reveals its symmetries; and calculation of the
asymptotic wave function distribution which shows that
directed waves fill space quite uniformly and are not
multifractal.

We thank Leon Balents, Sue Coppersmith, Matthew Fisher
and Onuttom Narayan for discussions and especially David
Huse for pointing out that the $g(\beta)$ spectrum
for our distribution is consistent with the data of
ref \cite{multifractal}.
H. Mathur thanks Matthew Fisher for hospitality at the ITP Santa Barbara
where this work was initiated. H. Mathur is supported
by NSF Grant DMR 98-04983 and an Alfred P Sloan Research
fellowship; the work at Santa Barbara was supported via
NSF grant PHY 94-07194. Yi-Kuo Yu was supported by the National Science
Foundation at the ALCOM Science and Technology Center
funded by Grant No. DMR 89-20147 and by startup funds from
Florida Atlantic University.

\end{multicols}


\begin{references}
\bibitem{onsite} For example, one could include a random on-site term
$V_{n}(x) \psi_{n}(x)$ in eq (1). However this term can be gauge
transformed into a hopping term leading to eq (1) and in the 
strong disorder limit the statistics of the induced hopping
term remain essentially white noise; see, for example, ref \cite{matthew}.
Another possibility is to discretise time as in ref \cite{saul}.
\end{references}
\end{document}